# *Dressed behavior of the quasiparticles lifetime in the unitary limit of two unconventional superconductors.*

P. Contreras[*,1] Dianela Osorio[1] and E. Yu. Beliayev[2]

[1] Department of Physics, University of the Andes, Mérida, Venezuela.
[2] Verkin Institute for Low Temperature Physics and Engineering, National Academy of Sciences of Ukraine, Kharkiv, Ukraine.
[*]Corresponding author (pcontreras@ula.ve)

**ABSTRACT**

We compare the quasiparticle lifetime behavior in the unitary limit of two unconventional superconductors dressed by non-magnetic impurity scattering to differentiate an anomalous functional behavior in its shape when the disorder concentration is changed in a triplet paired model with respect to the well behave singlet model. For singlet paired superconductors, the functional shape of the lifetime due to elastic scattering around the nodal regions does not change with the change of the disorder concentration, but for a triplet model with a tiny gap, an anomalous drop in shape is observed only when small values of disordering are added. We use a 2D tight-binding parametrization to study the reduced phase space of the first Brillouin zone, where the low energy scattering is restricted to the nodal/quasinodal regions for two irreducible representations of the crystal lattice.

## *1. Introduction*

The unitary regime in metallic alloys (in their normal and superconducting states) is defined as the limit where the elastic scattering of fermionic quasiparticles (normal, nodal/cuasinodal excitations) due to a non-magnetic disorder potential - $U_0$ is so strong that the fermionic mean-free path "$\ell$" becomes comparable to the inverse Fermi momentum "$k_F^{-1}$", or to the lattice parameter "a", (that is $\ell\, k_F \sim \ell\, a^{-1} \sim 1$), contrasting with the Born limit where the mean free path is much larger than "$k_F^{-1}$" and "a" (that is $\ell\, k_F \sim \ell\, a^{-1} \gg 1$) [1,2,3]. This physically means that the quasiparticles [4] (i. e. electrons and holes) in the unitary region have an ill-defined momentum $\boldsymbol{p} = \hbar \boldsymbol{k}$ between elastic collisions.

In the mathematical formalism, another signature of the unitary state in unconventional superconductors is the large resonance at zero frequency in the imaginary part "$\Im\,[\widetilde{\omega}(\omega + i\,0^+)]$" of the dressed energy "$\widetilde{\omega}(\omega + i\,0^+)$" [1,5]. On the other hand, in the normal state, the unitary limit gives a constant $\Im\,[\widetilde{\omega}(\omega + i\,0^+)]$ that corresponds to a constant lifetime ($\tau$) [1].

It is important to mention that the experimental approach to differentiate among the unitary, and the Born limits in unconventional superconductors, is to plot the ratio of the reduced temperature $T_c/T_{C0}$ as a function of the measured residual density of the states RDOS $N(0)/N_F$. This is made with the aid of the critical temperature $T_C$ reduction equation due to non-magnetic impurity scattering in unconventional superconductors [6]. Particularly, it has been shown that the unconventional superconductors $La_{2-x}Sr_xCuO_4$ (a high $T_C$ superconductor doped with strontium) [7,8,9,10], and $Sr_2RuO_4$ [11,12,13,14] (a triplet unconventional superconductor) are both in the unitary limit.

Therefore, unconventional superconductors can be defined as those with nodes/quasinodal regions around the Fermi surface, and with anisotropic order parameters (OP) with a spin paired dependence (singlet or triplet). The role of the non-magnetic elastic scattering mechanism is to break the superconducting state around those nodal/quasinodal regions creating a reduced fermion phase scattering space below $T_C$. For $Sr_2RuO_4$, this pair breaking mechanism is very sensible [15] because it contains Sr atoms in its unit cell, meanwhile, in $La_{2-x}Sr_xCuO_4$, the pair breaking mechanism is weaker since Sr is an external impurity added to the crystal lattice [9].

Following our previous works [16, 17], we use tight-binding OPs to model the 2D shape unitary behavior of the superconducting lifetime in two phenomenological models, for a Cuprate doped with Strontium and for Strontium Ruthenate. The energy (frequency) dependence of the superconducting lifetime comes in this case from the reduced phase space of scattering states and is due to the presence of the superconducting (SC) gap.

In the mathematical formalism - for unconventional superconductors, the imaginary part of the elastic scattering matrix $\widetilde{\omega}(\omega + i\,0^+)$, dressed by the nonmagnetic impurity potential $U_0$ defines the inverse of the dressed quasiparticles (electrons and holes) lifetime $\tau^{-1}(\omega)$ [3], for quasiparticles (electrons and holes), which are closed in the nodal regions by means of eq.

$$(2\tau)^{-1}(\omega) = \Im\left[\widetilde{\omega}(\omega + i\,0^+)\right] \quad [1]$$

The purpose of this work is to visualize the shape of the unitary quasiparticles lifetime $\tau^{-1}(\omega)$ evolving as a function of doping $\Gamma^+$, in the phase space where the low energy scattering is restricted to the nodal/quasinodal regions.

In order to calculate the shape of the function $\tau^{-1}(\omega)$, that differentiates singlet from triplet OPs, we present 2 tables in which we briefly describe the physical properties and input parameters of both models. Finally, we use a first neighbor tight-binding approximation (TB), which helps to visualize how atoms and impurities interact, since TB goes beyond the free energy quasiparticles model [4] and it is very suitable for 2D simulations with phenomenological parameters in terms of energy (in meV units). We also briefly compare our work with the shape of the lifetime in the case of 3D TB. We point out that at very low temperature the kinetic coefficients such as the electronic thermal conductivity and the ultrasound attenuation (under doping by non-magnetic impurities) can show the universal behavior in the unitary limit [4, 5, 14].

## 2. Models and method to visualize the shape of the dressed lifetime $(2\tau)^{-1}(\omega)$:

In Table 1, we summarize the models and approximations used in this work. The simulation is performed within the first Brillouin zone using a first neighbor TB approximation in both equations, the normal state energy and gap spectra with the respectively Fermi averages and the correspondent point symmetric lattice groups.

Table 1: Summary of the TB models and the group theoretical classification of the OPs used in this work.

| Reduced phase space of scattering states | Order parameter model | Electrons approximation | Metallic sheet of the Fermi surface | Order parameter (OP) symmetry | Compounds and crystals |
|---|---|---|---|---|---|
| Line nodes crossing the Fermi surface. | Scalapino first neighbor line nodes model [7] | first neighbors hoping with a symmetric tight binding electron-hole dispersion law | Located at the four corners of the 1st Brillouin zone | Scalar states $\phi(k)$ with even function of $k$ and even parity $l \in B_{1g}$ of the $D_{4h}$ point group. | Cuprate superconductors such as LSCO [8]: $La_{2-x}Sr_xCuO_4$ |
| Quasinodal, it does not intercept the $\gamma$ sheet of the Fermi surface. | Miyake & Narikiyo first neighbors tiny gap model for the FS $\gamma$ sheet [14] | idem | Located at the center of the 1st Brillouin zone | Vector states $d(k)$ with odd function of $k$ and odd parity $l \in E_{2u}$ of the $D_{4h}$ point group. | Triplet superconductors such as Strontium ruthenate: $Sr_2RuO_4$ [11] |

In this work we remark that the function $\widetilde{\omega}(\omega + i\,0^+)$ resembles an imaginary number with a self-consistent function as a coefficient accompanying the complex i, that is, $\widetilde{\omega}(\omega + i\,0^+) = \omega + i\,(2\tau)^{-1}(\widetilde{\omega})$. The quasiparticle lifetime for a single sheet on the Fermi Surface - $\tau^{-1}(\omega)$ is defined self-consistently. Following a realistic parametrization of $\widetilde{\omega}(\omega + i\,0^+)$ in terms of two parameters, c and $\Gamma^+$ introduced in [18], we use for the scattering lifetime the equation

$$(2\tau)^{-1}(\widetilde{\omega}(\omega), c, \Gamma^+) = \pi\,\Gamma^+ \frac{g(\widetilde{\omega})}{c^2 + g^2(\widetilde{\omega})},\ [2]$$

where $\widetilde{\omega}$ and $\omega$ are continuum variables in meV units, while c (dimensionless) and $\Gamma^+$ (meV) are the two phenomenological parameters, and the function $g(\widetilde{\omega}) = \langle \frac{\widetilde{\omega}}{\sqrt{\widetilde{\omega}^2 - |\Delta|^2(k_x, k_y)}} \rangle_{FS}$ depends on $\Delta$, the superconducting OP, averaged over the two TB Fermi sheets <...>$_{FS}$ using a method suitable to fit experimental low temperature data [19,20].

$\widetilde{\omega}(\omega + i\, 0^+)$ is a self-consistent nonlinear complex equation which only can be solved numerically, c and $\Gamma^+$, are the phenomenological parameters that describe the breaking of the superconducting pairs, driving the existence of normal state quasiparticles inside the superconductor with a finite lifetime strongly depending on the non-magnetic disorder. The parameter $c = 1/(\pi\, N_F\, U_0)$ is the inverse of the impurities potential $U_0$ and $N_F$ is the Fermi level DOS. The parameter $\Gamma^+ = n_{imp}/(\pi^2\, N_F)$ is proportional to the impurity concentration $n_{imp}$. Non-magnetic disorder in metals assumes N equal scatterers randomly distributed, but independent each other, it also assumes that on a macroscopic scale, the crystal is homogeneous [21].

The unitary limit corresponds to values $U_o \gg 1\ or\ c = 0$ in eq. 2

$$(2\tau_U)^{-1}(\widetilde{\omega}(\omega), c = 0, \Gamma^+) = \pi\, \Gamma^+ \frac{1}{g(\widetilde{\omega})}. \quad [3]$$

We use the TB equations and parameters according to table 2. In addition, we use for both the singlet and triplet gaps, the Fermi averages given by $g(\widetilde{\omega}) \sim \langle\widetilde{\omega}\rangle_{FS} \neq 0,\,,\ g_1(\widetilde{\omega}) \sim \langle\Delta(k_x, k_y)\rangle_{FS}= 0$, and $g_3(\widetilde{\omega}) \sim \langle\xi(k_x, k_y)\rangle_{FS}= 0$ [22].

Table 2: Summary of the equations of the energy dispersion and the OPs used in this work.

| Nodal behavior | equations | $(t, \epsilon_F, \Delta_0)$ values |
|---|---|---|
| First neighbors line nodes singlet OP | $\Delta(k_x, k_y) = \Delta_0\, \phi(k_x, k_y)$ $= [\cos(k_x a) - \cos(k_y a)]$ | (0, 0, 33.9 ) meV [23] |
| Fist neighbors quasi-nodal triplet OP | $\boldsymbol{\Delta}(k_x, k_y) = \Delta_0\, \boldsymbol{d^\gamma}(k_x, k_y)$ $= [(\sin(k_x a) + i \sin(k_y a)]\hat{\boldsymbol{z}}$ | (0, 0, 1.0 ) meV [17] |
| First neighbors anisotropic tight binding dispersion law | $\xi(k_x, k_y) = \epsilon_F + 2\, t\, [\cos(k_x a) + \cos(k_y a)]$ $\epsilon_F \sim 2\, t$ | (0.2, 0.4, 0) meV, centered at the points (±πa, ±πa). |
| First neighbors anisotropic tight binding dispersion law | $\xi(k_x, k_y) = \epsilon_F + 2\, t\, [\cos(k_x a) + \cos(k_y a)]$ $\epsilon_F \sim -\, t.$ | (0.4, - 0.4, 0) meV, centered at point (0, 0). |

## 3. Singlet case:

In the case of the singlet model, figure 1 visualizes the evolution of $(2\tau_U)^{-1}$ at c = 0 for the five values of $\Gamma^+$ in meV following eq. 3. We observe the smooth peak centered at zero residual frequencies $\gamma = \widetilde{\omega}(0)$ with a much smaller value of $\gamma$ for very dilute values of disorder concentration $\Gamma^+$. We see a monotonically flattening of the function for higher energies indicating a constant lifetime value in the unitary limit for the normal state for all values of $\Gamma^+$. We also observe that the value of $(2\tau_U)^{-1}$ in the normal state (above 33.9 meV) depends on the disorder concentration, given for an optimally doped disorder a value of $(2\tau_U)^{-1}$ = 0.40 meV, meanwhile the very dilute disorder case gives $(2\tau_U)^{-1}$ = 0.02 meV.

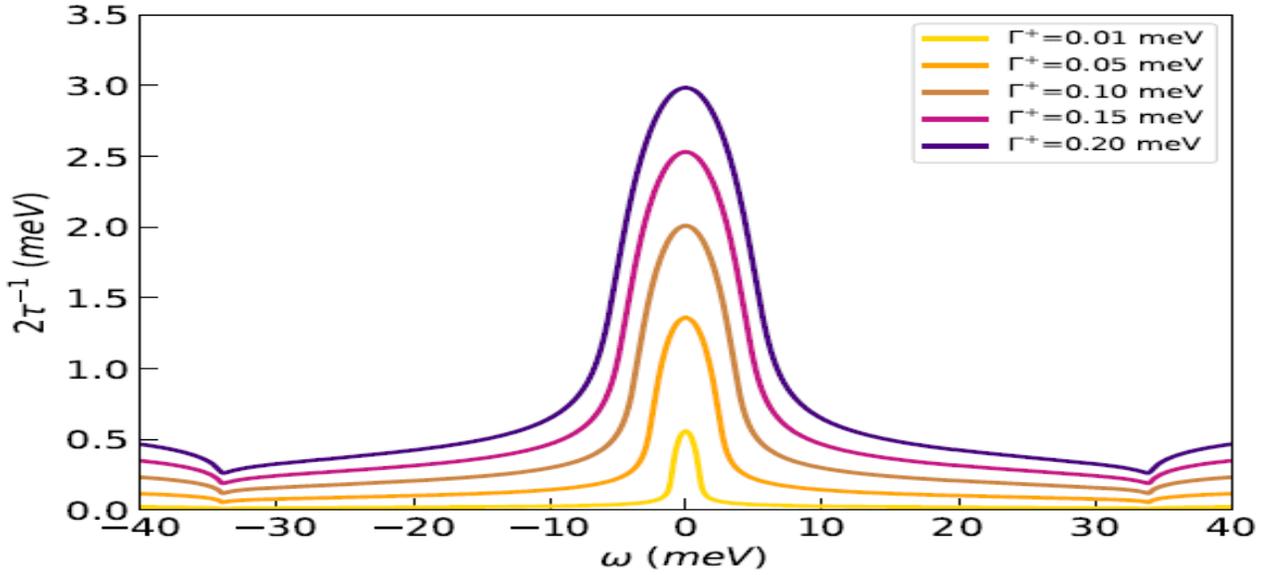

*Figure 1. Plot of the inverse scattering lifetime as a function of the concentration $\Gamma^+(meV)$ in the unitary limit for the singlet OP.*

We also observe from figure 1, a regular behavior of the shape of the quasiparticles lifetime function with the same shape for the five values of $\Gamma^+$. From the simulation, we also observe that for the lifetime in the normal state (above 33.9 meV) there is an increasing dependence on the value of $\Gamma^+$. At energies $\omega = \Delta_0 = \pm 33.9$ meV, we observe the transition from the superconducting to the normal state as a small abrupt change in the function's slope.

Finally, we notice that for a very small amount of impurity concentration ($\Gamma^+$ = 0.01 meV), the function $(2\tau_U)^{-1} \to 0$, that is, there is a significantly reduced phase scattering space for dressed electrons (yellow color curve) with a majority presence of supercurrent carriers. Also in figure 1, the yellow curve behavior can be considered almost constant for very low $\Gamma^+$ except around the small unitary resonance interval.

## 4. Triplet case:

In the case of the triplet model, figure 2 shows the evolution of $(2\tau_U)^{-1}$ according to eq. 3, and for the same values of $\Gamma^+$ as it was done in the singlet case we observe the smooth resonance centered at zero frequency for all values, with smaller values of residual zero energy $\gamma$ for very dilute values of disorder $\Gamma^+$. For the triplet case, the value $\gamma \sim \Delta_0 = 1$ meV for an optimal ($\Gamma^+ = 0.20$ meV) violet curve in figure 2, while for the singlet OP, we observe in figure 1 (violet curve) that $\gamma \sim \Delta_0/10 \approx 3$ meV for $\Gamma^+ = 0.20$ meV. In addition, the orange curve corresponds to dilute levels of disorder with $\Gamma^+ = 0.05$ meV. We also see for the triplet case a monotonically flattening of the function $(2\tau_U)^{-1}$ for higher energies indicating a constant lifetime value in the unitary limit for the normal state and for all values of $\Gamma^+$ with similar values as for the singlet case.

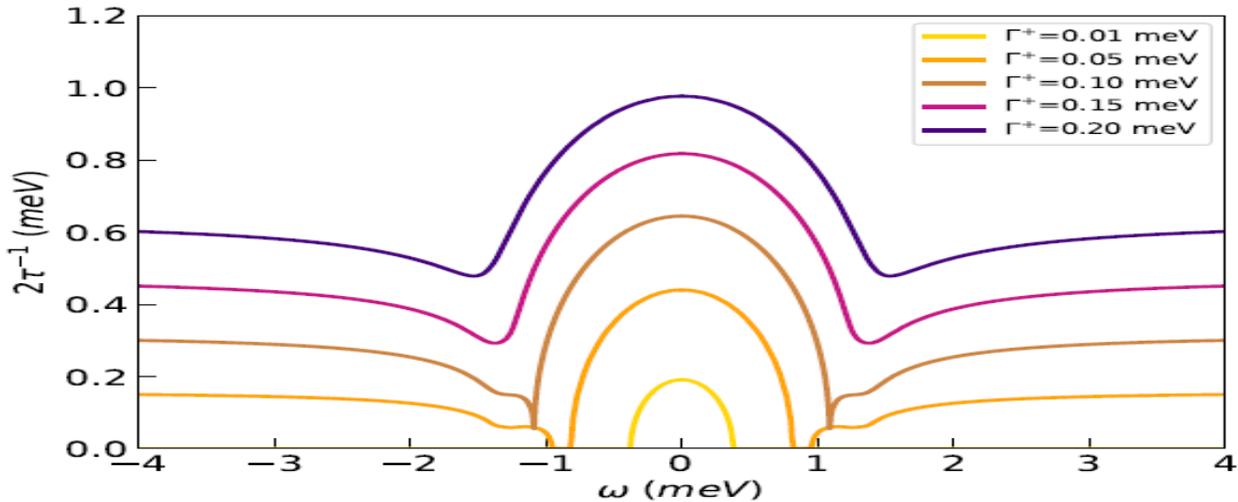

*Figure 2. Plot of the inverse scattering lifetime as a function of the concentration $\mathbf{\Gamma^+(meV)}$ in the unitary limit for the triplet OP.*

We observe that the value of $(2\tau_U)^{-1}$, in the normal state increases with disorder concentration. At energies $\omega \sim \Delta_0 = \pm 1.0$ meV, in figure 2, we do not observe for the triplet case the transition from the superconducting to the normal state, but instead starting from $\Gamma^+ = 0.10$ meV in the brown curve, a sharp minimum that evolves into a smooth minimum (displaced to higher frequencies) is shown in the violet and blue curves.

An anomalous behavior with $\omega \in (0.85, 1)$ meV, where $(2\tau_U)^{-1} = 0$ found previously [17] corresponds to the tiny gap effect in $(2\tau)^{-1}(\omega)$ [14] and has significance for a transition to a more complicated non-homogeneous BCS like unitary state with "frozen" quasinodal fermions.

## 5. Scattering in reduced phase space and physical kinetics

A classification for the two OPs based on physical kinetics [29] using an analysis of the reduced scattering space is as follows. Figure 3 shows on the left side a d wave singlet (a), and on the right a triplet (b) OP TB scaled scattering phase space. It follows from (a) that a regular shape behavior distinguishes only one region (1) with fermion quasiparticles and supercurrent carriers (no changing with disorder concentration Γ⁺) in the reduced scattering space with a maximum at zero residual frequency and the change in slope at the value of the transition, that is ω ≈ Δ₀,.

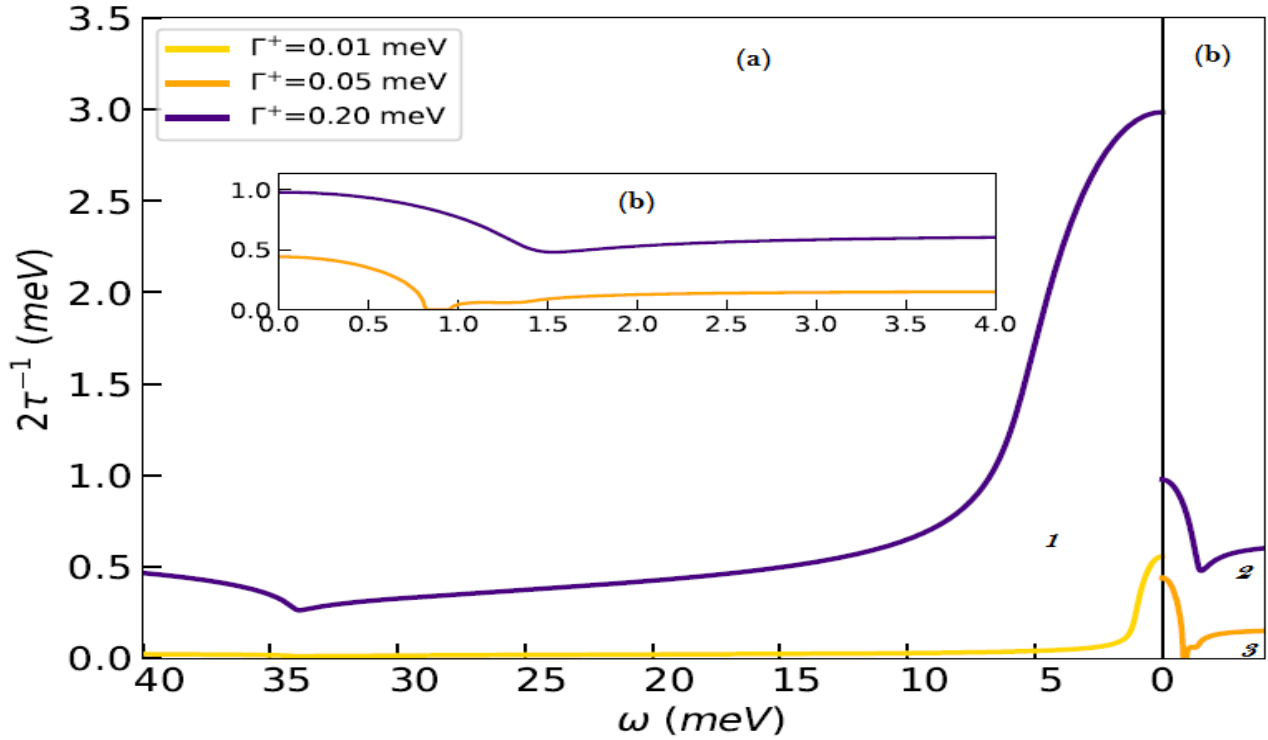

*Figure 3. Proper TB scaled homogenous unitary phases 1 and 2 ( violet and yellow curves), and an inhomogeneous unitary phase 3 (orange curve), in the reduced scattering space of the 2 OPs, notice that the reduced phase space in the case of the triplet OP (b) is smaller for any value of Γ⁺.*

Figure 3 also shows on the right side (b) and upper insert in (a), for a p wave triplet order parameter two well-distinguished behaviors (2) and (3) in reduced quasinodal scattering space depending on the non-magnetic disorder concentration Γ⁺. A regular shape behavior (2) for Γ⁺≥ 0.10 meV that distinguishes a first homogenous region in reduced phase space with a maximum at zero residual frequency γ and a smooth transition to the normal state that smears at ω ≈ 1.5 meV and the presence of both nodal fermions and supercarriers. There is also an irregular shape behavior (the Miyake Narikiyo tiny gap) in the region 3(b) for Γ⁺≲ 0.05 meV, that is, an interval with a maximum zero residual frequency γ ~ 0.5 meV, and totally frozen nodal quasiparticles, that is $(2\tau_U)^{-1} = 0$, close to ω ≈ Δ₀ = 1 meV. We call this an inhomogeneous unitary state since there won't be fermion transport carriers.

This difference between singlet and triplet OPs, induce us to think of an inhomogeneous reduced scattering phase 3 shown in (b) of figure 2 due to a weak disorder Γ⁺ in p- paired triplet superconductors such as strontium ruthenate, where one of the two compelling quasiparticles (the quasi-nodal fermions) disappear (orange curve insert (b) in figure 3), and only the coupled pairs condensate (supercarriers) remain in the unitary state where inhomogeneties play a crucial role since $\ell k_F$ ~ $\ell a^{-1}$ ~ 1.

Experiments concerning the scattering contribution for the different compelling quasiparticles in unconventional superconductors have been performed and analyzed for the case of the high Tc compound $Bi_2Sr_2CaCu_2O_{8+\delta}$ [24]. Moreover, the key role of disorder in $Sr_2RuO_4$ as a triplet paired state also is being intensively studied at this moment [25].

Additionally, we point out that in the works [26, 27, 28], it was found that $La_{2-x}Sr_xCuO_4$ has some peculiarities at very low concentrations of Sr (x = 0.001). The concentration of charge carriers did not correspond to the Neel temperature for this strontium concentration. These authors were able to show that, near this concentration, the distribution of strontium impurities over the sample is extremely inhomogeneous, and the sample tends to form metallic superconducting inclusions. Moreover, in a recent work [30] it has been observed that the charged carriers (electrons) disappear for a certain level of doping in the low temperature behavior of the compounds $Tl_2Ba_2CuO_{6+d}$ and $Bi_2Sr_2CuO_{6+d}$, we point out that this could be due to the tiny MN gap [14, 17] gap observed in figures 2 and 3.

To finalize, we add a few words about the dimensionality of the scattering space(2D/3D) and the shape of $\tau^{-1}(\omega)$ following figure 2 in [32]. For 3 D models that take into account the $k_z$ component, we know that the geometrical behavior of $\tau^{-1}(\omega)$ will be more complicated (self-consistency is not taken into account in a 3D analysis). In general, we can show that in that case, the energy dependence of the superconducting lifetime comes from the reduced phase volume of scattering states because of the presence of SC gap, and the frequency dependence of $\tau^{-1}(\omega)$ is explained by the total 3D energy spectrum $\epsilon(k) = \sqrt{\xi_k^2 + \Delta_k^2}$. In 3D, isolated zeros of $\epsilon(k)$ at given k with a linear energy-momentum relation in the vicinity of the zeros gives rise to the linear low energy behavior, saddle points in $\epsilon(k)$ give rise to logarithmic Van Hove singularities, and local maxima at certain k give rise to a step like feature.

In 2D, we, however, do not observe 3D minimax geometrical effects in scattering space, but we can appreciate effects such as: 1. The sudden drop of the inverse lifetime value to zero, which is due to the appearance of the tiny gap in the spectrum that can be obtained numerically only by fully solving the self-consistent problem for a triplet OP in the unitary limit. 2. An almost constant scattering lifetime for small $\Gamma^+$ in a singlet OP for the unitary limit.

## 6. Conclusion

This work was aimed at comparing the behavior of the inverse of superconducting lifetime

$(2\tau)^{-1}(\omega)$ shape in the unitary limit, when the quasiparticles are dressed by a non-magnetic impurity potential, for the case of two nodal/quasinodal OPs, -the line (Scalapino [7]) and the tiny gap (Miyake and Narikiyo [14]) models, using a 2D anisotropic TB parametrization.

In sections 2, 3, 4, and 5, the behavior of the dressed function $(2\tau_U)^{-1}(\widetilde{\omega}(\omega), \Gamma^+)$ in the unitary limit was numerically computed self-consistently for five values non-magnetic disorder potential $U_0 \gg 1$, starting with very diluted to optimal $\Gamma^+$ disorder values ($\Gamma^+$ = 0.01, 0.05, 0.10, 0.15, 0.20) [meV].

The results were visualized in figures 1, 2, and 3, noticing the following remarks:

- There is a unique signature among the two types, the zero frecuency maximum of the function $(2\tau_U)^{-1}(\widetilde{\omega}(\omega), \Gamma^+)$ for all values of the concentration parameter $\Gamma^+$ in the unitary limit.
- For the lines nodes model, the shape of $(2\tau_U)^{-1}(\widetilde{\omega}(\omega), \Gamma^+)$ does not change for the five values of $\Gamma^+$. The lifetime is well-behaving, in the sense that there won´t be any changes in the shape of the function for very small or, instead, reasonable amounts of non-magnetic disorder. Both kinds of quasiparticles are present in the nodal scattering phase space regions, supercarriers (bosons) and the low energy nodal excitations (fermionic), however at very low $\Gamma^+$, mostly supercarriers will be present since $(2\tau_U)^{-1}(\widetilde{\omega}(\omega), \Gamma^+) \to 0$.
- For the triplet tiny gap (quasinodes) model, the shape of $(2\tau_U)^{-1}(\widetilde{\omega}(\omega), \Gamma^+)$ changes among the five values of $\Gamma^+$ used. The lifetime is well-behaving, for relatively high values of $\Gamma^+$ (=0.10, 0.15, 0.20) meV with both fermionic quasinodal and supercarriers in the reduced phase space, but if only a very small amount $\Gamma^+$ = 0.05 meV of non-magnetic dirt is added, the function $(2\tau_U)^{-1}(\widetilde{\omega}(\omega), \Gamma^+)$ becomes identical to zero for a tiny interval of energy as shown graphically from the simulation in figures 2 and 3. Therefore, there is a tiny gap in the energy interval $\in [0.7, 1.0]$ meV where only supercarriers (bosons) are present.
- This can be thought of as anomalous behavior in the triplet model, which is no observed in the singlet model, and it could be interpreted phenomenologically as an inhomogeneity effect when only a small amount of Sr atoms become a non-magnetic dirt ($\Gamma^+$ = 0.05 meV) and the superconductor is in the unitary limit (c=0 and $\ell k_F \sim \ell a^{-1} \sim 1$). Higher values of dirt ($\Gamma^+ \geq 0.10$ meV) will smear out the effect observed in figures 2 and 3, mixing fermionic quasinodal particles and supercarriers.
- Finally, a hypothesis proposed in [31] that the scattering lifetime in unconventional superconductors is almost constant has been shown valid in our numerical calculation for the yellow curve in figures 1, 2 and 3 at very low $\Gamma^+$ except around the small zero resonance interval.

We, therefore, conclude that the 2D TB parametrization allowed us to formally study and compare self-consistently (in the first B. z.) for the unitary limit in the reduced scattering space where the low energy scattering of classical quasiparticles [4, 33] is restricted to the nodal/quasinodal regions of these two OP symmetries. In the unitary regime the mean free path and the lattice parameter are comparable, that is, $\ell a^{-1} \sim 1$, and inhomogeneity and non-magnetic disorder studies are relevant.

## 7. *Acknowledgements*

This paper is dedicated to the 100[th] birthday of Prof. Moisey Isaakovich Kaganov (1921 – 2021) .

This research did not receive any specific grant from funding agencies in the public, commercial, or not-for-profit sectors.